\documentclass[prb,showpacs,showkeys,preprintnumbers,superscriptaddress,a4paper,twocolumn]{revtex4-1}
\usepackage[T1]{fontenc}
\usepackage[utf8]{inputenc}
\usepackage{graphicx}
\usepackage{lmodern}
\usepackage[english]{babel}
\usepackage{amsmath,amssymb}
\usepackage[colorlinks=true,linkcolor=red]{hyperref}
\usepackage[sort&compress]{natbib}
\usepackage{color}														%	remove from the final version
\usepackage{graphicx}

\begin{document}

\title{Competing Antiferromagnetic and Spin-Glass Phases in a Hollandite Structure}

\author{Y. Crespo}
\affiliation{The Abdus Salam ICTP, Strada Costiera 11, I-34151 Trieste, Italy}
\author{A. Andreanov}
\affiliation{The Abdus Salam ICTP, Strada Costiera 11, I-34151 Trieste, Italy}
\affiliation{Max-Planck-Institut f\"ur Physik komplexer Systeme, N\"othnitzer Str. 38, 01187 Dresden, Germany}
\author{N. Seriani}
\affiliation{The Abdus Salam ICTP, Strada Costiera 11, I-34151 Trieste, Italy}

\begin{abstract}
We introduce a simple lattice model with Ising spins to explain recent experimental results on spin freezing in a hollandite-type structure. We argue that geometrical frustration of the lattice in combination with nearest-neighbour antiferromagnetic (AFM) interactions is responsible for the appearance of a spin-glass phase in presence of disorder.  We investigate this system numerically using parallel tempering. The model reproduces the magnetic behaviour of oxides with hollandite structure, such as $\alpha-\text{MnO}_2$ and presents a rich phenomenology: in absence of disorder three types of ground states are possible, depending on the relative strength of the interactions, namely AFM ordered and two different disordered, macroscopically degenerate families of ground states. Remarkably, for sets of AFM couplings having an AFM ground state in the clean system, there exists a critical value of the disorder for which the ground state is replaced by a spin-glass phase while maintaining all couplings AFM. To the best of our knowledge this is the only existing model that presents this kind of transition with short-range AFM interactions. We argue that this model could be useful to understand the relation between AFM coupling, disorder and the appearance of a spin-glass phase.
\end{abstract}

\date{\today}

\pacs{
75.10.Hk 	%Classical spin models
75.10.Nr 	%Spin glass and other random models
75.50.Lk 	%Spin glasses and other random magnets
}
\keywords{antiferromagnetism, geometrical frustration, hollandite structure, Ising, manganese oxide, spin-glass}
\maketitle

\section{Introduction}

Spin-glasses are magnetic phases where disorder and frustration suppress any simple ordered patterns, like ferro- or antiferromagnetic states, and have instead a spin-freezing transition into an amorphous \emph{glassy} ordered  state at low temperatures (or other control parameters). Such glassy states feature many interesting and unusual properties, like power-law correlations in the absence of any broken symmetry~\cite{de1983eigenvalues} and non-trivial long-time behaviour. The free-energy landscape of such systems is very rough and has many metastable states that are separated by high barriers~\cite{mezard1987spin}.

The two crucial ingredients necessary to produce a spin-glass are disorder and frustration. The canonical representatives of spin-glasses, magnetic alloys, have an oscillating long-range Ruderman-Kittel-Kasuya-Yosida (RKKY) spin-spin interaction~\cite{mydosh1993spin}. The standard Edwards-Anderson model~\cite{edwards1975theory} has quenched random nearest-neighbour couplings of both signs which mimics the frustrating effect of the longer-ranged RKKY interaction. In both cases, as well as in many others, it is the interaction, which is both disordered and frustrating (\textit{i.e.} mixture of ferromagnetic (FM) and antiferromagnetic (AFM) couplings) that is responsible for appearance of a spin-glass. There is also a large class of materials where frustration has geometrical origin:  the combination of the AFM interactions and geometry of the lattice suppresses the natural AFM order~\cite{henley2010the} and makes the system extremely susceptible to perturbations. In this case often even a small disorder in the coupling strengths, that does not change their AFM character, is enough to obtain a spin-glass~\cite{saunders2007spin,andreanov2010spin,shinaoka2011spin}. Consider now a geometrical frustrated system with nearest-neighbour AFM couplings of different strengths. Can the AFM ground state still survive? If so, is it possible to obtain a spin-glass phase by tuning the disorder and maintaining the AFM interactions? Do we need to introduce a small amount or a large disorder to break the AFM ground state?
 
% Can this constraints be relaxed ? That is can there be a spin glass in a system with frustrated couplings but has no disorder, 
% or in a system where the interaction alone is non-frustrating ? Some models with alternating but non-random distribution
% of couplings were studied~\cite{villain1977spin}. 

In this paper we introduce a model which addresses these points and extends the previously studied models of spin-glass transition in presence of geometrical frustration. The interactions in the model are geometrically frustrated with all the couplings being AFM but of different strengths. The natural AFM order is not always suppressed in the clean case or by disorder; a critical amount of disorder is required to suppress it. We can tune the strength of the geometrical frustration by changing the relative values of the couplings and obtain three  different types of ground states: one is non-degenerate and AFM ordered; the other two are exponentially degenerate and disordered. The disorder in the coupling strengths which leaves the couplings AFM, puts all these ground states in competition and leads to the spin-glass state. This provides an example of a system where the geometry forces different ground states to compete and there is a direct transition from an AFM state to a spin-glass phase by tuning the disorder, while having only AFM couplings. To the best of our knowledge this is the only existing model that presents this kind of transition with short-range, nearest-neighbour AFM interactions.

This model has been developed to explain the magnetic behaviour of manganese (Mn) oxides with the hollandite structure, such as $\alpha$-MnO$_2$~\cite{DeGuzman,Suib}, K$_{1.5}$(H$_3$O)$_x$Mn$_8$O$_{16}$~\cite{satoenoki}, Ba$_{1.2}$Mn$_8$O$_{16}$~\cite{Ishiwata}, \textit{ etc.}. As we will show below, experimentally these materials have a rich magnetic phenomenology whose origin is poorly understood. In particular it is not clear if the magnetic properties can be explained by considering only the bulk properties like the effects of doping and frustration or if surface effects have to be taken into account. Therefore a theoretical work is needed for a better understanding of the origin of the magnetic behaviour of these materials. This kind of studies will open the venue for a systematic approach to tune and optimise the properties of these compounds.

\begin{figure}[!tbp]
\begin{center}
\includegraphics[width=\columnwidth]{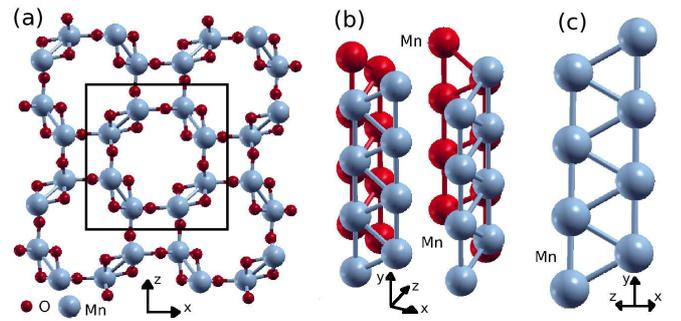}
\caption{(Colour online) Different views of the hollandite structure for the case of $\alpha$-$\text{MnO}_2$ compound~\cite{luo2009spin,luo2010tuning} (a) the xz plane containing both Mn and oxygen (O) atoms (b) panoramic view of the channel with only Mn atoms (c) Structural triangles composed by magnetic atoms that play an important role in setting the magnetic properties of the oxide.}
\label{fig:MnO2-lattice}
\end{center}
\end{figure}

Hollandite-type Mn oxides are nanoporous materials~\cite{DeGuzman,Suib} that are interesting on its own for their large number of applications, as ionic conductor, as catalysts for oxygen reduction and oxygen evolution reactions, in lithium-air batteries~\cite{bruce1, girishkumar, parkreview}, in supercapacitors~\cite{lifengou}, for the energy extraction from salinity differences~\cite{cui} and as a water oxidation catalyst~\cite{boppana}. The large lateral size of the channels (see Fig.~\ref{fig:MnO2-lattice} (a)), of the order of 0.46 nm, make it possible for some big cations such as K$^+$, Na$^+$, Ba$^{2+}$ and H$_3$O$^+$ to be introduced during material synthesis~\cite{umek, luo2009spin,luo2010tuning, satoenoki, Ishiwata}, thus opening the possibility to tune the magnetic, chemical and physical properties by cation doping. The magnetic nature of this compound originates from the presence of the manganese ions, which have localised magnetic moments due to the open shell of $d$ electronic states~\cite{satoenoki, umek}. These properties together with the presence of magnetic frustration present because of the existence of structural triangles (see Fig.~\ref{fig:MnO2-lattice} (c)), are at the origin of the rich variety of magnetic ordered phases that have been identified experimentally in these materials.

For example $\alpha$-MnO$_2$ in absence of doping elements in the channels, \textit{i.e.} in absence of disorder, has an AFM phase at low temperatures~\cite{Yamamoto, langong}. Luo \textit{et al.} have investigated the dependence of the magnetic phase on the doping of potassium ions ~\cite{luo2009spin,luo2010tuning}, and have shown that above a critical concentration of K$^+$ ions, a difference in the magnetic susceptibilities between zero-field-cooled and field-cooled samples is observed which is interpreted as the onset of the 
spin-glass phase. When all the sites available for K insertion became full, the AFM behaviour was recovered. The spin-glass phase was previously observed in the case of the KMn$_8$O$_{16}$ material \cite{Suib} with the same behaviour of the magnetic susceptibilities. Lan and co-workers proposed a surface effect to be responsible for the spin-glass behaviour in sodium-doped $\alpha$-MnO$_2$ nanorods~\cite{langong}. They proposed that, at low temperature, the core of the rod is in an AFM state while the surface spins contribute to a net magnetic moment of the specimen~\cite{langong}. Their explanation thus relies on the high surface area present in the channels of the material more than on the doping. The magnetic behaviour can become even more complex by including more than one doping species. For example Sato \textit{et al.} have investigated K$_{1.5}$(H$_3$O)$_x$Mn$_8$O$_{16}$(0<x<0.5), and found three magnetic transitions by varying the temperature~\cite{satoenoki}. They explained their results on the basis of a FM phase existing at intermediate temperature, and a ground state with helical magnetism appearing below 20 K. 

Due to the complex magnetism present in these materials, it is important, as a first step, to disentangle the contributions of bulk effects (doping and frustration) and surface effects in order to gain a better understanding and control of the properties of these compounds. The model introduced in this paper is able to reproduce the essential features of these systems, in particular the transition taking place at increasing doping/disorder between an AFM ground state and a spin-glass phase, on the basis of bulk properties only.

The paper is organised as follows: we introduce the model describing the magnetic properties of manganese oxide in Sec.~\ref{sec:model} and study its properties in the clean limit in Sec.~\ref{sec:clean}. Next we discuss the mechanism which generates the spin-glass phase in the doped samples and present the results of numerical simulations in support of our model in Sec.~\ref{sec:disorder}. Conclusions are presented at the end. 

\section{Model}
\label{sec:model}

The materials of interest for this study have a complicated lattice structure known as the hollandite lattice. This structure consists of two octahedra joined at the edges to form the wall of $2\times2$ channels (see Fig.~\ref{fig:MnO2-lattice}), and belongs to a family of crystals which differ from each other only by the lateral size of the channels, like ramsdellite ($1\times2$), romanechite ($2\times3$) and todorokite ($3\times3$)~\cite{DeGuzman,thackeray, thackeray2, shenzerger, devaraj}. In the case of manganese oxides each octahedron is formed by a $\text{MnO}_6$ unit. 

\begin{figure}[!htbp]
\begin{center}
\includegraphics[width=\columnwidth]{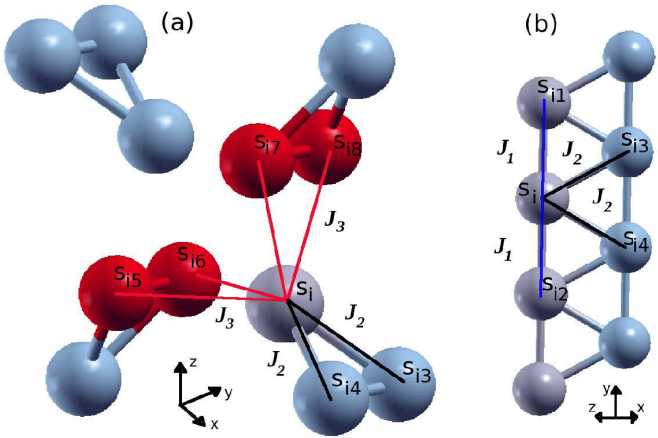}
\caption{(Colour online) The graphical representation of the Hamiltonian~\eqref{eqn:H}. Links of different colour represent the three couplings $J_1$ (blue), $J_2$ (black) and $J_3$ (red).}
\label{fig:MnO2-interactions}
\end{center}
\end{figure}

The magnetism is due to the interaction of the magnetic moments localised on the manganese ions, which interact with each other through oxygen-mediated superexchange~\cite{satoenoki, umek, kanamori, good1, good2}. We consider the simplest possible model compatible with the lattice structure and place classical Ising spins on Mn sites of the hollandite lattice. Therefore we get a lattice of spins which has $8$ spins per unit cell as shown in Fig.~\ref{fig:MnO2-lattice}. The Hamiltonian reads:
\begin{gather}
\mathcal{H} = \sum\limits_{<ij>_1} J_1^{ij} s_is_j + \sum\limits_{<ij>_2} J_2^{ij} s_is_j + \sum\limits_{<ij>_3} J_3^{ij} s_is_j,
\label{eqn:H}
\end{gather}
where $<ij>_k$ denotes three different groups of nearest-neighbours of a spin. The partitioning of the neighbours and the corresponding  couplings $J_1$, $J_2$, $J_3$ are detailed in Fig.~\ref{fig:MnO2-interactions}. Such division and choice of the three different coupling constants are dictated by the the structure of the material: all these three classes of nearest neighbours are quite close to the atom: the distance to atoms of the $1^{st}$ group is $\sim$2.86 \AA, while the atoms of the  $2^{nd}$ and 3$^{rd}$ groups are at distances of 2.91 \AA ~and 3.44 \AA, respectively. The choice of Ising spins is the simplest possible and, as we will show, it already reproduces the main experimentally observed features.

Despite the simplicity of the Hamiltonian~\eqref{eqn:H} it admits many different phases depending on relative strengths and signs of the couplings $J_k$ for $k=1,2,3$. This is due to the complicated geometrical structure of the hollandite lattice. If all the couplings are FM, the phase diagram only has paramagnetic and FM phases. Since the undoped compound is known to have an AFM ordering experimentally~\cite{Yamamoto}, we assume AFM couplings $J_k$, $k=1,2,3$ throughout the rest of the paper. This choice corresponds to the maximum frustration of the interactions and reproduces experimentally observed features.

\section{Clean system}
\label{sec:clean}

\begin{figure}[!htbp]
\begin{center}
\includegraphics[width=\columnwidth]{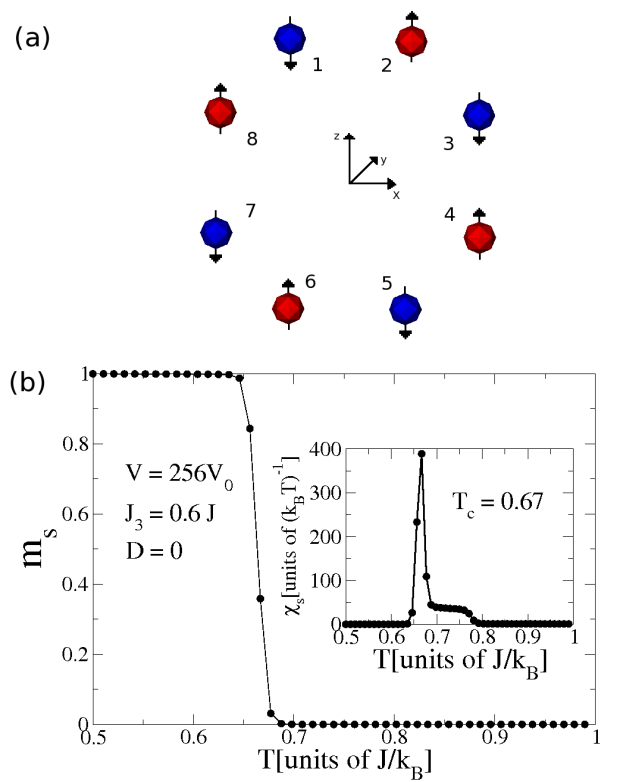}
\caption{(a) (Colour online) C-type antiferromagnetic (C-AFM) ground state on the hollandite lattice at $T=0$ for $2J_3 > J$. There is an AFM alignment in the clockwise/counterclockwise direction. (b) Staggered magnetisation $m_s$ as implied by the ground state of (a) as a function of temperature for the clean system (D=0), $N=2048$ spins, volume $V=256V_0$, where $V_0$ is the volume of the unit cell. The couplings are equal to $J_1=J_2=J=1$ and $J_3=0.6J$. The inset: the staggered susceptibility $\chi_s$ as a function of temperature. The sharp peak signalises the transition from the paramagnetic phase to N\'eel order.}
\label{fig:model-clean-af}
\end{center}
\end{figure}

We start by studying the clean limit where the strengths of all of the couplings $J_k$ $k=1,2,3$ are constant and do not fluctuate in space (no disorder). Experiments indicate a transition to an AFM phase~\cite{Yamamoto} at low enough temperatures. This behaviour is reproduced in our model if we set the couplings to $J_1=J_2=J$ and $J/2 < J_3 \leq J$ where we have taken $J=1$ for all the calculations. We assume $J_1 = J_2$ since the structural triangles formed by Mn atoms are almost equilateral~\cite{luo2009spin} (see Fig.~\ref{fig:MnO2-interactions}). With this choice of  the interaction strengths there is a unique ground state with spins ordered in a C-type AFM (C-AFM) state as shown in Fig.~\ref{fig:model-clean-af} (a).

The C-type AFM order (C-AFM) is composed of FM chains which are coupled antiferromagnetically~\cite{C_AFM_Kajimoto} (see Fig.~\ref{fig:model-clean-af}). We now give a simple minimisation argument in favour of the C-AFM ordering for this particular choice of the coupling strengths. As can be seen from Fig.~\ref{fig:MnO2-interactions} every Mn atom $s_i$ interacts with $8$ other Mn atoms $s_{ik}, k=1,\ldots,8$. The coupling strengths for these $8$ atoms are as follows: two with $J_1$ ($s_{i1}$,$s_{i2}$), two with $J_2$ ($s_{i3}$,$s_{i4}$) and four with $J_3$ ($s_{im}, m=5,6,7,8$). Therefore as long as $2J_3 > J$, there is a unique state that minimises the total energy and it has $s_{im}=-s_{ij}=-s_i$, $j=1,2$. This arrangement of spins does not minimise the energy for the links in the $y$ direction $s_{i1}=s_{i2}=s_i$ where spins are aligned ferromagnetically and the AFM links are unsatisfied. However this is exactly compensated by the interaction energy with the spins $s_{i3}=s_{i4}=-s_i$. The ground state average energy per spin is $<E_0> = -2J_3$. This ordered C-AFM configuration can be described as follows: if one numerates the $8$ spins in the unit cell in the clockwise (or anticlockwise) direction $s_n$, $n=1,\dots,8$ (see Fig.~\ref{fig:model-clean-af} (a)) then $s_n=(-1)^n$. This unit cell can be replicated on all the three directions implying that in the direction of the channels, along the $y$ axis, the spins are aligned ferromagnetically (see Fig.~\ref{fig:model-clean-af} (a)).

We have checked the validity of the above argument by performing classical Monte-Carlo simulations \cite{MC} of a clean system with $N=2048$ Ising spins and a simulation cell of volume $V=256V_0$ where $V_0$ is the volume of the unit cell composed by $8$ spins. We assumed periodic boundary conditions and we have chosen $J_3 = 0.6J$, $J_1=J_2=J$. The standard Metropolis \cite{MC} rule with single spin-flip updates was used. We computed the staggered magnetisation which we defined as $m_s = M_s/N$ with:
\begin{equation}
M_s = \sum_{i=1}^N (-1)^i s_i.
\end{equation}
where the site index $i$ is the same as in the minimisation argument presented above and in Fig.~\ref{fig:MnO2-interactions} ($i_{1-8}$). The values of $m_s$ computed using the Monte-Carlo for various values of temperature are shown in Fig~\ref{fig:model-clean-af} (b). Clearly $m_s = 0$  for high temperatures and $m_s\to 1$ for temperatures close to zero indicating the AFM order. We also computed the staggered susceptibility $\chi_s = dm_s/dh_s |_{h_s=0} = (<M_s^2> -<M_s>^2)/k_BT$ where $k_B$ is the Boltzmann constant and $T$ the temperature. As one can see in the inset of Fig.~\ref{fig:model-clean-af} (b) the staggered susceptibility has a sharp peak at the same temperature where $m_s$ becomes non-zero. The critical temperature $T_c = 0.67 J/k_B$  is estimated from the divergence of $\chi_s$ (see inset of Fig.~\ref{fig:model-clean-af} (b)). 

\begin{figure}[!tbp]
\begin{center}
\includegraphics[width=\columnwidth]{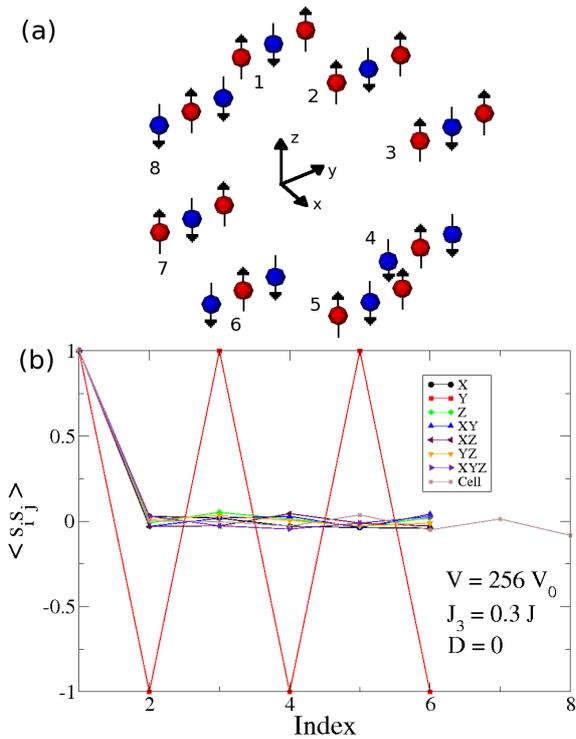}
\caption{(a) (Colour online) Single unit cell replicated in the $y$ direction. The spin configuration corresponds to one of the possible  ground states with energy per spin $<E_0>=-J$. The spins are always aligned antiferromagnetically in the $y$ direction for these ground states, while there is no order in the perpendicular direction. (b) The spin-spin correlations in the geometrically frustrated phase for the $x,y,z,xy,xz,yz,xyz$ directions and inside the unit cell. Here the index is defined as the number of times you have to move in each direction to find $s_j$ {\it e.g.} in the $xy$ direction index=2 means that $s_j$ is located at a distance of $\sqrt{(2L_x)^2 + (2L_y)^2}$ where $L_x$ and $L_y$ are the length of the unit cell in the $x$ and $y$ direction respectively. In the case of ``Cell'' the index runs over the 8 atoms of the unit cell.}
\label{fig:model-clean-gf-corr}
\end{center}
\end{figure}

For $2J_3 < J$ the situation is completely different: the C-AFM phase is suppressed and the staggered magnetisation is always zero: $m_s=0$ as confirmed by Monte-Carlo simulations. Instead there is an extensive number of spin configurations that minimise the total energy as we show below. The ground states in this phase can be rationalised as follows: if $2J_3 < J$ then the coupling to the nearest-neighbours in the $1^{st}$ and the $2^{nd}$ groups is more important than interaction with the $3^{rd}$ group of nearest-neighbours, unlike the case $2J_3 > J$. To better explain the spin configurations that minimise the total energy let's concentrate on one spin $s_i$ and assume $s_i=1$ (see Fig. \ref{fig:MnO2-interactions}). Since $2J_3 < J$, $s_i$ interacts the strongest with the nearest-neighbour spins from the  $1^{st}$ and the $2^{nd}$ groups. Therefore we select $s_{i1}=s_{i2}=-s_i=-1$, and this creates an AFM order along the chain of Mn atoms in the $y$ direction. Next we have to fix the spins $s_{i3}$ and $s_{i4}$, we notice that each of them interacts equally with two spins, one up and one down, in the chain containing $s_i$ (see Fig. \ref{fig:MnO2-lattice} (c)), and therefore the interaction energy will be the same irrespective of the orientation of the spins $s_{i3}$ and $s_{i4}$. Then the energy is reduced by minimising the interaction between $s_{i3}$ and $s_{i4}$, that is setting $s_{i3}=-s_{i4}$ and creating a second AFM chain along the $y$ direction that contains the $s_{i3}$ spin. In this configuration the interaction of $s_i$ with the $s_{i3}$ and $s_{i4}$ cancels out. Repeating the same reasoning for the other spins, we minimise the energy for the nearest neighbours of the $1^{st}$ and the $2^{nd}$ groups in all the four Mn triangular chains (see Fig. \ref{fig:MnO2-lattice} (b)). The interaction of $s_i$ of with spins of the $3^{rd}$ nearest-neighbour group also cancels out since $s_{i5}=-s_{i6}$ and $s_{i7}=-s_{i8}$ and the average total energy per spin is $<E_0>=-J$. To summarise this picture one can think that the ground state is composed by a collection of one-dimensional Mn-Mn chains each having AFM order of spins. However the chains are completely uncorrelated between them and arranged in such a way to form the channels of the hollandite structure (see Fig.~\ref{fig:model-clean-gf-corr} (a)). That is there is a perfect AFM order in $y$ direction, but no order in the transverse direction. If we consider a single unit cell replicated in the $y$ direction then we will have $2^8$ possibles configurations with the same energy per spin $<E_0>=-J$. Since each chain is completely uncorrelated to the others, for an increasing number of cells in the $x$ and $z$ directions, $N_\text{cell}^{xz}$,the number of ground states is growing as $2^{8 N_\text{cell}^{xz}}$, \textit{i.e.} exponentially.

The Monte-Carlo results confirm this picture: for $2J_3 < J$ the staggered magnetisation is always zero and the staggered susceptibility  does not display any sharp features down to $T=0$. Such behaviour is typical for systems where the combination of lattice geometry and interactions suppresses the natural AFM order. At low temperatures, instead, they enter a phase known as collective paramagnet~\cite{henley2010the} with power-law/exponential correlations depending on the properties of the lattice (bipartite/non-bipartite). We have computed the spin-spin correlation functions along different axes of the unit cell as shown in Fig.~\ref{fig:model-clean-gf-corr} (b). The simulation was performed for the same system as presented in Fig.~\ref{fig:model-clean-af} (b) but with $J_3=0.3J$. Unlike other geometrically frustrated systems there is a clear AFM alignment of spins along the $y$ direction and a (fast) exponential decay as we deviate from this direction ($x,z,xy,xz,yz,xyz$) and also with the 8 spins of the unit cell. This confirms the picture developed above, based on simple energy minimisation argument.

For $2J_3 = J$ the system is even more degenerate: both families of the ground states introduced above have equal energies, allowing also for configurations that locally mix both solutions, making it difficult to put forward a minimisation argument for this case. The system has an exponential number of ground states. It has no order at low temperatures and the AFM alignment along the $y$ direction is suppressed. We believe that this lends support to the view that the system might be in a classical spin liquid phase.

To summarise: in the clean limit, with only AFM couplings there are three types of ground states. The one with C-AFM order ($2J_3 > J$), a second one where there is a perfect anticorrelation along the one-dimensional Mn-Mn chains and no order is present between chains ($2J_3 < J$) and a third ground state consisting in a mixture of the previous two ($2J_3 = J$).

\section{Disordered system}
\label{sec:disorder}

We now turn to the disordered case and study what effect the disorder has on the physical properties of the system described by the Hamiltonian~\eqref{eqn:H}. A spin-glass phase was experimentally observed in different compounds, for example $\text{KMn}_8\text{O}_{16}$ material~\cite{Suib} and in the case of $\text{Na}^+$ and $\text{K}^+$ doped $\alpha$-$\text{MnO}_2$~\cite{langong,luo2009spin,luo2010tuning}. Usually spin-glass behaviour is associated with the presence of disorder and frustration in the system~\cite{villain1977spin}. As discussed above, the clean system has geometrical frustration. We suggest the following mechanism to explain the experimental results: as the dopant ions penetrate into the channels of the hollandite structure \cite{umek}, they locally modify the compound and therefore also the magnetic interactions between the spins, i.e. they generate fluctuations in the AFM couplings $J_k$, $k=1,2,3$. In principle, this could happen through different mechanisms, such as doping-induced strain or changes in the local electronic structure through charge donation.

In the case of $\alpha$-$\text{MnO}_2$ it has been experimentally observed that lattice strains induced by doping are minor~\cite{luo2009spin}, while it seems more probable that the charge donated by the doping elements modifies the oxidation state of manganese, producing a local mixture of Mn$^{+3}$ and Mn$^{4+}$~\cite{DeGuzman,luo2009spin,Shen}. This local change in the electronic properties induces fluctuations in the strength of magnetic spin-spin interaction, breaking the symmetry of the system (all Mn atoms are equivalent in the clean case). In our model we mimic these fluctuations by introducing continuous disorder in the couplings $J_k$. We would like to stress that we assume fluctuations of the couplings around their clean values so that the couplings remain AFM. The presence of quenched fluctuations in the couplings $J_k$ is crucial for the appearance of the spin-glass phase~\cite{saunders2007spin,andreanov2010spin,shinaoka2011spin}. Since the compounds investigated in the experiments have an AFM order at low temperatures in the clean limit and the appearance of the spin-glass requires doping to be higher than some critical value, this places us in the $2J_3 > J$ part of the phase diagram of our model.
%Therefore we begin to introduce disorder on this part of the phase diagram.
 
We propose the following mechanism for the appearance of the spin-glass in the doped compounds: randomness in the AFM couplings $J_k$, $k=1..3$ produces local regions of the lattice where the condition $2J_3 > J$ is violated and that local patch has the C-AFM ordering locally disfavoured. The presence of such patches frustrates the system, and generates the spin-glass phase. Such mechanism implies that the spin-glass can only appear for sufficiently strong fluctuations in the strengths of the couplings: one needs sufficiently big number of the patches. Therefore we expect the critical strength of the disorder to be an increasing function of the distance to the point $2J_3 = J$. Since the amplitude of the fluctuations (disorder) depends on the doping, we need a sufficiently large doping to produce the spin-glass. This is exactly what is seen in experiments: Luo \textit{et al.}~\cite{luo2009spin,luo2010tuning} have found that a critical amount of K$^+$ doping in $\alpha$-K$_x$MnO$_2$ is needed to see the spin-glass phase.

\begin{figure}[!tb]
%\centering{\includegraphics[width=7.0cm,angle=270]{Ms.smalld.eps}}
%\centering{\includegraphics[width=\columnwidth]{Ms.smalld.eps}}
\centering{\includegraphics[width=\columnwidth]{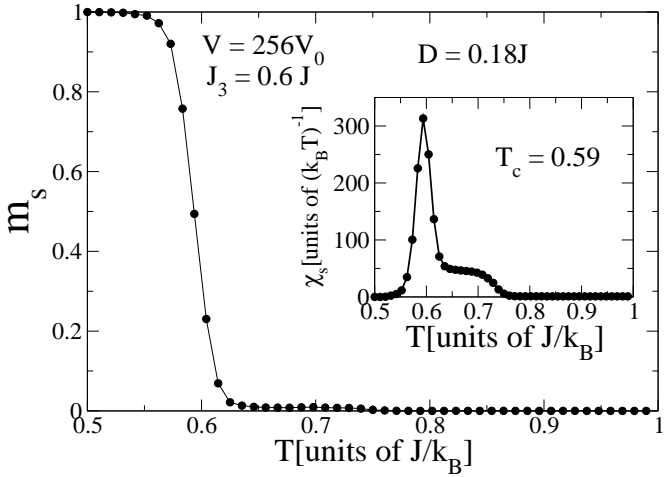}}
\caption{Staggered magnetisation $m_s$ as a function of temperature for the same system as in Fig.~\ref{fig:model-clean-af} except  now a small amount of disorder $D = 0.18 J$ drawn from the box distribution is present in the AFM couplings. This disorder strength is not enough to destroy the C-AFM phase. Inset: staggered susceptibility $\chi_s$ as a function of temperature for the same system and the same disorder strength.}
\label{fig:Ms_sm_fig}
%\hspace{5mm}
\end{figure}

We have performed numerical simulations of the model with disorder to confirm our ideas. The AFM couplings $J_{ij}=J_{ji}=J_k + \Delta$ are different for each link connecting a pair of atoms $i,j=1..N$. The disorder $\Delta$ is drawn from the box distribution $-D < \Delta < D$ where $D$ sets the scale for the strength of the quenched fluctuations. Glassy systems are notoriously difficult to simulate at low temperatures and we used parallel tempering~\cite{marinari1992simulated,hukushima1996exchange,hartmann2002optimization} to equilibrate the samples at low temperatures. We used up to $N_s = 200$ samples to compute the disorder average. In every case we have tested that convergence to the average was achieved.

We first show the results for small amounts of disorder $D$ where the C-AFM phase is still present. In Fig.~\ref{fig:Ms_sm_fig} we show the plot of the staggered magnetisation vs. temperature for $D=0.18 J$. The parameters of the system are the same as those of the clean case ($D=0$, see Fig.~\ref{fig:model-clean-af}):$J_1 = J_2 = J$ and $J_3 = 0.6 J$,  $N=2048$ spins, $V=256V_0$ where $V_0$ is the volume of the unit cell. We see that this amount of disorder is not enough to destroy the C-AFM phase but the critical temperature $T_c = 0.59 J/k_B$ extracted from the position of the sharp peak of the staggered susceptibility (see Fig.~\ref{fig:Ms_sm_fig}) is reduced in comparison with the clean case where $T_c=0.67 J/k_B$.   

Upon further increase of the strength of the disorder the C-AFM phase is completely suppressed at $D = 0.4$ and the spin-glass phase appears instead at low temperatures. To characterize the paramagnetic to spin-glass transition we have studied the Binder cumulant~\cite{Binder,Bhatt} which is defined as:
\begin{equation}
G = \frac{1}{2}\left[3 - \frac{\left<q^4\right>}{\left<q^2\right>^2} \right],
\label{Bq_eq}
\end{equation}
where $\left<\cdots\right>$ denote both thermal average for a given disorder and the average over the disorder, and $q$ is the overlap between two independent replicas of the system ($s_i^1,s_i^2$) with the same realisation of disorder:
\begin{equation}
q =  \frac{1}{N} \sum_{i=1}^N s_i^1 s_i^2.
\end{equation}
The Binder cumulant $G$ is a dimensionless parameter that goes to zero for high temperatures and it is of order one in the spin-glass phase and has the finite size scaling form~\cite{Bhatt} close to the transition:
\begin{equation}
G = g\left[ V^{1/3\nu}(T-T_c)\right] 
\label{eq:SBq_eq}
\end{equation}
where $V$ is the volume of the sample and $T_c$ is the critical temperature. These properties make the Binder parameter $G$ a useful tool to study spin-glass phase transitions and it has been widely used for this purpose~\cite{Kawashima,Ballesteros,Herrero,Thomas}, as all curves of $G$ generated for different system sizes will intersect at $T_c$. As a complementary parameter we have also computed the overlap distribution $P(q)$ that has the finite size scaling form~\cite{Bhatt}:
\begin{equation}
P(q) = V^{\beta/3\nu}p(q)\left[ qV^{\beta/3\nu},V^{1/3\nu}(T-T_c))\right] 
\label{eq:SPq_eq}
\end{equation}
where $\nu$ and $\beta$ are the critical exponents which are obtained by fitting the data for different system sizes to a single master curve. 

\begin{figure}[!tb]
%\centering {\includegraphics[width=7.0cm,angle=270]{Bp.eps}}
%\centering {\includegraphics[width=\columnwidth]{Bp.eps}}
\centering {\includegraphics[width=\columnwidth]{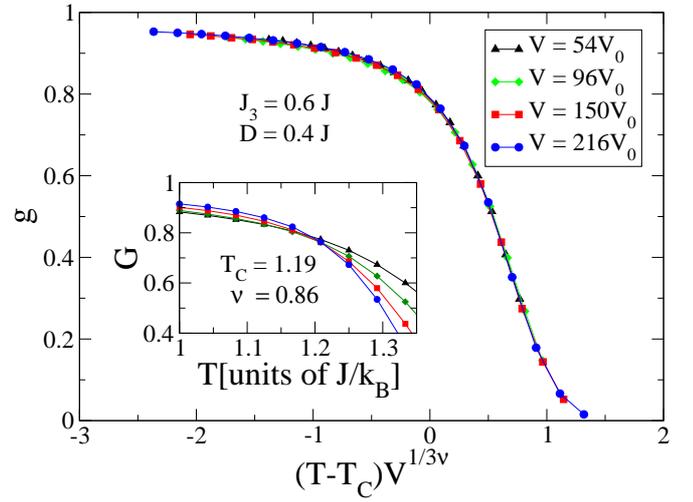}}
\caption{(Colour online) Scaled Binder cumulant as defined in Eq.~\eqref{eq:SBq_eq} for four system sizes $V=54V_0$, $V=96V_0$, $V=150V_0$, $V=256V_0$  where $V_0$ is the volume of the unit cell, the interactions $J_1=J_2=J$ and $J_3=0.6J$ where considered with a disorder of $D=0.4$. Inset: Binder cumulant as defined in Eq.~\eqref{Bq_eq} as a function of temperature.}
\label{fig:Bp_fig}
%\hspace{5mm}
\end{figure}

The parallel tempering simulations were performed for systems with AFM couplings $J_1=J_2=J$, $J_3=0.6J$ and disorder $D=0.4J$. The system sizes are $V=54V_0$,$V=96V_0$, $V=150V_0$, $V=256V_0$. In all the four cases the C-AFM phase was suppressed in favour of a spin-glass phase. The inset of Fig.~\ref{fig:Bp_fig} shows the Binder cumulant as a function of temperature for the four system sizes. There is a clear crossing of the Binder cumulant curves indicating the spin-glass transition near $T_c=1.19 J/k_B$ for $D=0.4J$.

\begin{figure}[!tb]
%\centering{\includegraphics[width=7.0cm,,angle=270]{Pq.eps}}
%\centering{\includegraphics[width=0.9\columnwidth]{Pq.eps}}
\centering{\includegraphics[width=0.9\columnwidth]{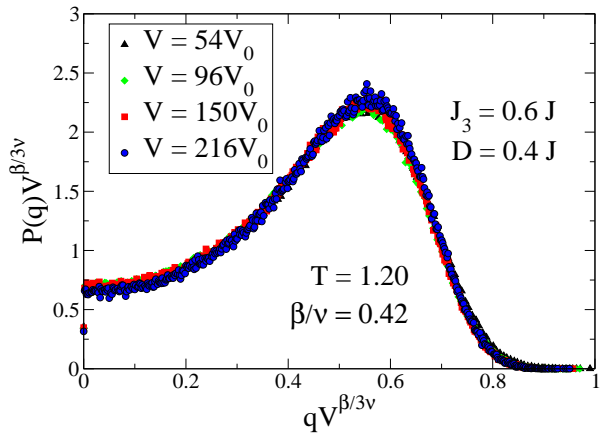}}
\caption{(Colour online) Scaling for the probability distribution $P(q)$ at $T=1.20 J/k_B$ near the critical temperature $T_c = 1.19 J/k_B$ using Eq.~\eqref{eq:SPq_eq} for the same system of Fig.~\ref{fig:Bp_fig}.}
\label{fig:Pq_fig}
%\hspace{5mm}
\end{figure}

Using this estimate for $T_c$ we collapsed all the data onto a single master curve according to Eq.~\eqref{eq:SBq_eq} as shown in Fig.~\ref{fig:Bp_fig} whit the critical exponent $\nu=0.86$. We have also studied the overlap distribution $P(q)$ for the four system sizes at the temperature $T=1.20 J/k_B$, close to the critical temperature $T_c=1.19 J/k_B$ (see Fig.~\ref{fig:Pq_fig}). Rescaling the $P(q)$ curves for different system sizes according to Eq.~\eqref{eq:SPq_eq} a good collapse on a master curve is obtained for the ratio of the critical exponents $\beta/\nu = 0.42$ from witch we get $\beta = 0.36$.

The values $\beta=0.36$ and $\nu=0.86$ are very different from the values of the critical exponents for the spin-glass transition in $d=3$: $\beta=0.77(5)$, $\nu=2.45(15)$~(see Ref.~\cite{hasenbusch2008critical} and references therein). These exponents are known to be hard to evaluate precisely due to large finite-size effects~\cite{hasenbusch2008critical}. As there are no other factors that could explain the discrepancy we attribute it to the smallness of the system sizes studied. Understanding better this discrepancy and studying in detail the dependence of the critical exponents on the system size requires to simulate much larger system sizes which is beyond the scope of this paper.

\section{Conclusions}
\label{sec:conclusions}

We have introduced a simple classical Ising spin model with nearest-neighbour antiferromagnetic (AFM) interactions and continuous disorder. This model explains recent experimental results on the magnetic properties of some manganese oxides with hollandite crystal structure. The studied system is geometrically frustrated due to the existence of structural triangles (see Fig.~\ref{fig:MnO2-lattice} (c)) in the hollandite lattice. The parallel tempering Monte-Carlo simulations performed for this system with clean and disordered AFM coupling strengths and different system sizes, reveal that, despite the simplicity of the Hamiltonian~ Eq.~\eqref{eqn:H}) (only nearest neighbour interactions), it admits different phases depending on relative strengths of the AFM couplings due to the complicated geometry of the hollandite lattice. In fact there are three types of ground states in the clean case: a non-degenerated one with a C-type AFM (C-AFM) order, composed of FM chains coupled with an AFM order between them (see Fig.~\ref{fig:model-clean-af}). The other two types of ground states are exponentially degenerated and disordered ground states. Depending on the strengths of the interactions a small amount of disorder in the couplings does not suppress the C-AFM order, but  merely reduces the critical temperature at which the paramagnet to C-AFM transition occurs (see Fig.~\ref{fig:Ms_sm_fig}). When the disorder becomes greater than some critical value, the geometry forces different ground states to compete and there is a direct transition from the C-AFM state to a spin-glass by tuning the disorder (at fixed temperature), while preserving AFM couplings. Such a transition was confirmed in numerically by looking at the Binder cumulant (see Fig.~\ref{fig:Bp_fig}). To the best of our knowledge this is the only existing model that presents this kind of transition with nearest-neighbours AF interactions.

This model represents a good starting point to understand the magnetic phases appearing in hollandite oxides. It reproduces the main features of the experimental results both in clean and doped cases. This model shows that bulk effects (frustration and doping) are sufficient to obtain the spin-glass in compounds with hollandite lattice. It would be interesting to check our predictions experimentally and to see how adequate the proposed mechanism for generating the spin-glass order is. It would also be useful to see the convergence of the critical exponents to their standard $d=3$ values, although this requires considerable amount of computing time. Finally the predictive power of our model could be enhanced by using couplings determined by {\it ab-initio} calculations.

\textbf{Acknowledgment.} We would like to thank Erio Tosatti, Peter Young, Roderick M\"ossner, Matteo Palassini and Yasir Iqbal for useful discussions. Computational resources were provided by CINECA, through the ISCRA-C project AFSGMNO2, and by The Abdus Salam ICTP. 

\bibliography{toy_model}

%merlin.mbs apsrev4-1.bst 2010-07-25 4.21a (PWD, AO, DPC) hacked
%Control: key (0)
%Control: author (8) initials jnrlst
%Control: editor formatted (1) identically to author
%Control: production of article title (-1) disabled
%Control: page (0) single
%Control: year (1) truncated
%Control: production of eprint (0) enabled
\begin{thebibliography}{44}%
\makeatletter
\providecommand \@ifxundefined [1]{%
 \@ifx{#1\undefined}
}%
\providecommand \@ifnum [1]{%
 \ifnum #1\expandafter \@firstoftwo
 \else \expandafter \@secondoftwo
 \fi
}%
\providecommand \@ifx [1]{%
 \ifx #1\expandafter \@firstoftwo
 \else \expandafter \@secondoftwo
 \fi
}%
\providecommand \natexlab [1]{#1}%
\providecommand \enquote  [1]{``#1''}%
\providecommand \bibnamefont  [1]{#1}%
\providecommand \bibfnamefont [1]{#1}%
\providecommand \citenamefont [1]{#1}%
\providecommand \href@noop [0]{\@secondoftwo}%
\providecommand \href [0]{\begingroup \@sanitize@url \@href}%
\providecommand \@href[1]{\@@startlink{#1}\@@href}%
\providecommand \@@href[1]{\endgroup#1\@@endlink}%
\providecommand \@sanitize@url [0]{\catcode `\\12\catcode `\$12\catcode
  `\&12\catcode `\#12\catcode `\^12\catcode `\_12\catcode `\%12\relax}%
\providecommand \@@startlink[1]{}%
\providecommand \@@endlink[0]{}%
\providecommand \url  [0]{\begingroup\@sanitize@url \@url }%
\providecommand \@url [1]{\endgroup\@href {#1}{\urlprefix }}%
\providecommand \urlprefix  [0]{URL }%
\providecommand \Eprint [0]{\href }%
\providecommand \doibase [0]{http://dx.doi.org/}%
\providecommand \selectlanguage [0]{\@gobble}%
\providecommand \bibinfo  [0]{\@secondoftwo}%
\providecommand \bibfield  [0]{\@secondoftwo}%
\providecommand \translation [1]{[#1]}%
\providecommand \BibitemOpen [0]{}%
\providecommand \bibitemStop [0]{}%
\providecommand \bibitemNoStop [0]{.\EOS\space}%
\providecommand \EOS [0]{\spacefactor3000\relax}%
\providecommand \BibitemShut  [1]{\csname bibitem#1\endcsname}%
\let\auto@bib@innerbib\@empty
%</preamble>
\bibitem [{\citenamefont {De~Dominicis}\ and\ \citenamefont
  {Kondor}(1983)}]{de1983eigenvalues}%
  \BibitemOpen
  \bibfield  {author} {\bibinfo {author} {\bibfnamefont {C.}~\bibnamefont
  {De~Dominicis}}\ and\ \bibinfo {author} {\bibfnamefont {I.}~\bibnamefont
  {Kondor}},\ }\href {\doibase 10.1103/PhysRevB.27.606} {\bibfield  {journal}
  {\bibinfo  {journal} {Phys. Rev. B}\ }\textbf {\bibinfo {volume} {27}},\
  \bibinfo {pages} {606} (\bibinfo {year} {1983})}\BibitemShut {NoStop}%
\bibitem [{\citenamefont {Mezard}\ \emph {et~al.}(1987)\citenamefont {Mezard},
  \citenamefont {Parisi},\ and\ \citenamefont {Virasoro}}]{mezard1987spin}%
  \BibitemOpen
  \bibfield  {author} {\bibinfo {author} {\bibfnamefont {M.}~\bibnamefont
  {Mezard}}, \bibinfo {author} {\bibfnamefont {G.}~\bibnamefont {Parisi}}, \
  and\ \bibinfo {author} {\bibfnamefont {M.~A.}\ \bibnamefont {Virasoro}},\
  }\href@noop {} {\emph {\bibinfo {title} {Spin glass theory and beyond}}},\
  Vol.~\bibinfo {volume} {9}\ (\bibinfo  {publisher} {World scientific
  Singapore},\ \bibinfo {year} {1987})\BibitemShut {NoStop}%
\bibitem [{\citenamefont {Mydosh}(1993)}]{mydosh1993spin}%
  \BibitemOpen
  \bibfield  {author} {\bibinfo {author} {\bibfnamefont {J.~A.}\ \bibnamefont
  {Mydosh}},\ }\href@noop {} {\emph {\bibinfo {title} {Spin Glasses: An
  Experimental Introduction}}}\ (\bibinfo  {publisher} {Taylor and Francis
  Inc.},\ \bibinfo {year} {1993})\BibitemShut {NoStop}%
\bibitem [{\citenamefont {Edwards}\ and\ \citenamefont
  {Anderson}(1975)}]{edwards1975theory}%
  \BibitemOpen
  \bibfield  {author} {\bibinfo {author} {\bibfnamefont {S.~F.}\ \bibnamefont
  {Edwards}}\ and\ \bibinfo {author} {\bibfnamefont {P.~W.}\ \bibnamefont
  {Anderson}},\ }\href {http://stacks.iop.org/0305-4608/5/i=5/a=017} {\bibfield
   {journal} {\bibinfo  {journal} {J. Phys. F: Metal Phys.}\ }\textbf {\bibinfo
  {volume} {5}},\ \bibinfo {pages} {965} (\bibinfo {year} {1975})}\BibitemShut
  {NoStop}%
\bibitem [{\citenamefont {L.}(2010)}]{henley2010the}%
  \BibitemOpen
  \bibfield  {author} {\bibinfo {author} {\bibfnamefont {H.~C.}\ \bibnamefont
  {L.}},\ }\href {\doibase 10.1146/annurev-conmatphys-070909-104138} {\bibfield
   {journal} {\bibinfo  {journal} {Annu. Rev. Condens. Matter Phys.}\ }\textbf
  {\bibinfo {volume} {1}},\ \bibinfo {pages} {179} (\bibinfo {year}
  {2010})}\BibitemShut {NoStop}%
\bibitem [{\citenamefont {Saunders}\ and\ \citenamefont
  {Chalker}(2007)}]{saunders2007spin}%
  \BibitemOpen
  \bibfield  {author} {\bibinfo {author} {\bibfnamefont {T.~E.}\ \bibnamefont
  {Saunders}}\ and\ \bibinfo {author} {\bibfnamefont {J.~T.}\ \bibnamefont
  {Chalker}},\ }\href {\doibase 10.1103/PhysRevLett.98.157201} {\bibfield
  {journal} {\bibinfo  {journal} {Phys. Rev. Lett.}\ }\textbf {\bibinfo
  {volume} {98}},\ \bibinfo {pages} {157201} (\bibinfo {year}
  {2007})}\BibitemShut {NoStop}%
\bibitem [{\citenamefont {Andreanov}\ \emph {et~al.}(2010)\citenamefont
  {Andreanov}, \citenamefont {Chalker}, \citenamefont {Saunders},\ and\
  \citenamefont {Sherrington}}]{andreanov2010spin}%
  \BibitemOpen
  \bibfield  {author} {\bibinfo {author} {\bibfnamefont {A.}~\bibnamefont
  {Andreanov}}, \bibinfo {author} {\bibfnamefont {J.~T.}\ \bibnamefont
  {Chalker}}, \bibinfo {author} {\bibfnamefont {T.~E.}\ \bibnamefont
  {Saunders}}, \ and\ \bibinfo {author} {\bibfnamefont {D.}~\bibnamefont
  {Sherrington}},\ }\href {\doibase 10.1103/PhysRevB.81.014406} {\bibfield
  {journal} {\bibinfo  {journal} {Phys. Rev. B}\ }\textbf {\bibinfo {volume}
  {81}},\ \bibinfo {pages} {014406} (\bibinfo {year} {2010})}\BibitemShut
  {NoStop}%
\bibitem [{\citenamefont {Shinaoka}\ \emph {et~al.}(2011)\citenamefont
  {Shinaoka}, \citenamefont {Tomita},\ and\ \citenamefont
  {Motome}}]{shinaoka2011spin}%
  \BibitemOpen
  \bibfield  {author} {\bibinfo {author} {\bibfnamefont {H.}~\bibnamefont
  {Shinaoka}}, \bibinfo {author} {\bibfnamefont {Y.}~\bibnamefont {Tomita}}, \
  and\ \bibinfo {author} {\bibfnamefont {Y.}~\bibnamefont {Motome}},\ }\href
  {\doibase 10.1103/PhysRevLett.107.047204} {\bibfield  {journal} {\bibinfo
  {journal} {Phys. Rev. Lett.}\ }\textbf {\bibinfo {volume} {107}},\ \bibinfo
  {pages} {047204} (\bibinfo {year} {2011})}\BibitemShut {NoStop}%
\bibitem [{\citenamefont {DeGuzman}\ \emph {et~al.}(1994)\citenamefont
  {DeGuzman}, \citenamefont {Shen}, \citenamefont {Neth}, \citenamefont {Suib},
  \citenamefont {O'Young}, \citenamefont {Levine},\ and\ \citenamefont
  {Newsam}}]{DeGuzman}%
  \BibitemOpen
  \bibfield  {author} {\bibinfo {author} {\bibfnamefont {R.~N.}\ \bibnamefont
  {DeGuzman}}, \bibinfo {author} {\bibfnamefont {Y.}~\bibnamefont {Shen}},
  \bibinfo {author} {\bibfnamefont {E.~J.}\ \bibnamefont {Neth}}, \bibinfo
  {author} {\bibfnamefont {S.~L.}\ \bibnamefont {Suib}}, \bibinfo {author}
  {\bibfnamefont {C.}~\bibnamefont {O'Young}}, \bibinfo {author} {\bibfnamefont
  {S.}~\bibnamefont {Levine}}, \ and\ \bibinfo {author} {\bibfnamefont {J.~M.}\
  \bibnamefont {Newsam}},\ }\href@noop {} {\bibfield  {journal} {\bibinfo
  {journal} {Chem. Mater.}\ }\textbf {\bibinfo {volume} {6}},\ \bibinfo {pages}
  {815} (\bibinfo {year} {1994})}\BibitemShut {NoStop}%
\bibitem [{\citenamefont {Suib}\ and\ \citenamefont {Iton}(1994)}]{Suib}%
  \BibitemOpen
  \bibfield  {author} {\bibinfo {author} {\bibfnamefont {S.~L.}\ \bibnamefont
  {Suib}}\ and\ \bibinfo {author} {\bibfnamefont {L.~E.}\ \bibnamefont
  {Iton}},\ }\href@noop {} {\bibfield  {journal} {\bibinfo  {journal} {Chem.
  Mater.}\ }\textbf {\bibinfo {volume} {6}},\ \bibinfo {pages} {429} (\bibinfo
  {year} {1994})}\BibitemShut {NoStop}%
\bibitem [{\citenamefont {Sato}\ \emph {et~al.}(1999)\citenamefont {Sato},
  \citenamefont {Enoki}, \citenamefont {Yamaura},\ and\ \citenamefont
  {Yamamoto}}]{satoenoki}%
  \BibitemOpen
  \bibfield  {author} {\bibinfo {author} {\bibfnamefont {H.}~\bibnamefont
  {Sato}}, \bibinfo {author} {\bibfnamefont {T.}~\bibnamefont {Enoki}},
  \bibinfo {author} {\bibfnamefont {J.-I.}\ \bibnamefont {Yamaura}}, \ and\
  \bibinfo {author} {\bibfnamefont {N.}~\bibnamefont {Yamamoto}},\ }\href
  {\doibase 10.1103/PhysRevB.59.12836} {\bibfield  {journal} {\bibinfo
  {journal} {Phys. Rev. B}\ }\textbf {\bibinfo {volume} {59}},\ \bibinfo
  {pages} {12836} (\bibinfo {year} {1999})}\BibitemShut {NoStop}%
\bibitem [{\citenamefont {Ishiwata}\ \emph {et~al.}(2006)\citenamefont
  {Ishiwata}, \citenamefont {Bos}, \citenamefont {Huang},\ and\ \citenamefont
  {Cava}}]{Ishiwata}%
  \BibitemOpen
  \bibfield  {author} {\bibinfo {author} {\bibfnamefont {S.}~\bibnamefont
  {Ishiwata}}, \bibinfo {author} {\bibfnamefont {J.}~\bibnamefont {Bos}},
  \bibinfo {author} {\bibfnamefont {Q.}~\bibnamefont {Huang}}, \ and\ \bibinfo
  {author} {\bibfnamefont {R.}~\bibnamefont {Cava}},\ }\href@noop {} {\bibfield
   {journal} {\bibinfo  {journal} {J. Phys.: Condens. Matter}\ }\textbf
  {\bibinfo {volume} {18}},\ \bibinfo {pages} {3745} (\bibinfo {year}
  {2006})}\BibitemShut {NoStop}%
\bibitem [{\citenamefont {Bruce}\ \emph {et~al.}(2012)\citenamefont {Bruce},
  \citenamefont {Freunberger}, \citenamefont {Hardwick},\ and\ \citenamefont
  {Tarascon}}]{bruce1}%
  \BibitemOpen
  \bibfield  {author} {\bibinfo {author} {\bibfnamefont {P.~G.}\ \bibnamefont
  {Bruce}}, \bibinfo {author} {\bibfnamefont {S.~A.}\ \bibnamefont
  {Freunberger}}, \bibinfo {author} {\bibfnamefont {L.~J.}\ \bibnamefont
  {Hardwick}}, \ and\ \bibinfo {author} {\bibfnamefont {J.-M.}\ \bibnamefont
  {Tarascon}},\ }\href@noop {} {\bibfield  {journal} {\bibinfo  {journal}
  {Nature Mater.}\ }\textbf {\bibinfo {volume} {11}},\ \bibinfo {pages} {19}
  (\bibinfo {year} {2012})}\BibitemShut {NoStop}%
\bibitem [{\citenamefont {Girishkumar}\ \emph {et~al.}(2010)\citenamefont
  {Girishkumar}, \citenamefont {McCloskey}, \citenamefont {Luntz},
  \citenamefont {Swanson},\ and\ \citenamefont {Wilcke}}]{girishkumar}%
  \BibitemOpen
  \bibfield  {author} {\bibinfo {author} {\bibfnamefont {G.}~\bibnamefont
  {Girishkumar}}, \bibinfo {author} {\bibfnamefont {B.}~\bibnamefont
  {McCloskey}}, \bibinfo {author} {\bibfnamefont {A.~C.}\ \bibnamefont
  {Luntz}}, \bibinfo {author} {\bibfnamefont {S.}~\bibnamefont {Swanson}}, \
  and\ \bibinfo {author} {\bibfnamefont {W.}~\bibnamefont {Wilcke}},\
  }\href@noop {} {\bibfield  {journal} {\bibinfo  {journal} {J. Phys. Chem.
  Lett.}\ }\textbf {\bibinfo {volume} {1}},\ \bibinfo {pages} {2193} (\bibinfo
  {year} {2010})}\BibitemShut {NoStop}%
\bibitem [{\citenamefont {Song}\ \emph {et~al.}(2011)\citenamefont {Song},
  \citenamefont {Park}, \citenamefont {Alamgir}, \citenamefont {Cho},\ and\
  \citenamefont {Liu}}]{parkreview}%
  \BibitemOpen
  \bibfield  {author} {\bibinfo {author} {\bibfnamefont {M.-K.}\ \bibnamefont
  {Song}}, \bibinfo {author} {\bibfnamefont {S.}~\bibnamefont {Park}}, \bibinfo
  {author} {\bibfnamefont {F.~M.}\ \bibnamefont {Alamgir}}, \bibinfo {author}
  {\bibfnamefont {J.}~\bibnamefont {Cho}}, \ and\ \bibinfo {author}
  {\bibfnamefont {M.}~\bibnamefont {Liu}},\ }\href@noop {} {\bibfield
  {journal} {\bibinfo  {journal} {Mater. Sci. Eng. R: Reports}\ }\textbf
  {\bibinfo {volume} {72}},\ \bibinfo {pages} {203} (\bibinfo {year}
  {2011})}\BibitemShut {NoStop}%
\bibitem [{\citenamefont {Li}\ \emph {et~al.}(2010)\citenamefont {Li},
  \citenamefont {Feng}, \citenamefont {Ou}, \citenamefont {Wu}, \citenamefont
  {Fu},\ and\ \citenamefont {Tong}}]{lifengou}%
  \BibitemOpen
  \bibfield  {author} {\bibinfo {author} {\bibfnamefont {G.-R.}\ \bibnamefont
  {Li}}, \bibinfo {author} {\bibfnamefont {Z.-P.}\ \bibnamefont {Feng}},
  \bibinfo {author} {\bibfnamefont {Y.-N.}\ \bibnamefont {Ou}}, \bibinfo
  {author} {\bibfnamefont {D.}~\bibnamefont {Wu}}, \bibinfo {author}
  {\bibfnamefont {R.}~\bibnamefont {Fu}}, \ and\ \bibinfo {author}
  {\bibfnamefont {Y.-X.}\ \bibnamefont {Tong}},\ }\href@noop {} {\bibfield
  {journal} {\bibinfo  {journal} {Langmuir}\ }\textbf {\bibinfo {volume}
  {26}},\ \bibinfo {pages} {2209} (\bibinfo {year} {2010})}\BibitemShut
  {NoStop}%
\bibitem [{\citenamefont {La~Mantia}\ \emph {et~al.}(2011)\citenamefont
  {La~Mantia}, \citenamefont {Pasta}, \citenamefont {Deshazer}, \citenamefont
  {Logan},\ and\ \citenamefont {Cui}}]{cui}%
  \BibitemOpen
  \bibfield  {author} {\bibinfo {author} {\bibfnamefont {F.}~\bibnamefont
  {La~Mantia}}, \bibinfo {author} {\bibfnamefont {M.}~\bibnamefont {Pasta}},
  \bibinfo {author} {\bibfnamefont {H.~D.}\ \bibnamefont {Deshazer}}, \bibinfo
  {author} {\bibfnamefont {B.~E.}\ \bibnamefont {Logan}}, \ and\ \bibinfo
  {author} {\bibfnamefont {Y.}~\bibnamefont {Cui}},\ }\href@noop {} {\bibfield
  {journal} {\bibinfo  {journal} {Nano. Lett.}\ }\textbf {\bibinfo {volume}
  {11}},\ \bibinfo {pages} {1810} (\bibinfo {year} {2011})}\BibitemShut
  {NoStop}%
\bibitem [{\citenamefont {Boppana}\ and\ \citenamefont {Jiao}(2011)}]{boppana}%
  \BibitemOpen
  \bibfield  {author} {\bibinfo {author} {\bibfnamefont {V.~B.~R.}\
  \bibnamefont {Boppana}}\ and\ \bibinfo {author} {\bibfnamefont
  {F.}~\bibnamefont {Jiao}},\ }\href@noop {} {\bibfield  {journal} {\bibinfo
  {journal} {Chem. Commun.}\ }\textbf {\bibinfo {volume} {47}},\ \bibinfo
  {pages} {8973} (\bibinfo {year} {2011})}\BibitemShut {NoStop}%
\bibitem [{\citenamefont {Umek}\ \emph {et~al.}(2009)\citenamefont {Umek},
  \citenamefont {Gloter}, \citenamefont {Pregelj}, \citenamefont {Dominko},
  \citenamefont {Jagodi{\v{c}}}, \citenamefont {Jagli{\v{c}}i{\'c}},
  \citenamefont {Zimina}, \citenamefont {Brzhezinskaya}, \citenamefont
  {Poto{\v{c}}nik}, \citenamefont {Filipi{\v{c}}}, \citenamefont {Levstik},\
  and\ \citenamefont {Ar{\v{c}}on}}]{umek}%
  \BibitemOpen
  \bibfield  {author} {\bibinfo {author} {\bibfnamefont {P.}~\bibnamefont
  {Umek}}, \bibinfo {author} {\bibfnamefont {A.}~\bibnamefont {Gloter}},
  \bibinfo {author} {\bibfnamefont {M.}~\bibnamefont {Pregelj}}, \bibinfo
  {author} {\bibfnamefont {R.}~\bibnamefont {Dominko}}, \bibinfo {author}
  {\bibfnamefont {M.}~\bibnamefont {Jagodi{\v{c}}}}, \bibinfo {author}
  {\bibfnamefont {Z.}~\bibnamefont {Jagli{\v{c}}i{\'c}}}, \bibinfo {author}
  {\bibfnamefont {A.}~\bibnamefont {Zimina}}, \bibinfo {author} {\bibfnamefont
  {M.}~\bibnamefont {Brzhezinskaya}}, \bibinfo {author} {\bibfnamefont
  {A.}~\bibnamefont {Poto{\v{c}}nik}}, \bibinfo {author} {\bibfnamefont
  {C.}~\bibnamefont {Filipi{\v{c}}}}, \bibinfo {author} {\bibfnamefont
  {A.}~\bibnamefont {Levstik}}, \ and\ \bibinfo {author} {\bibfnamefont
  {D.}~\bibnamefont {Ar{\v{c}}on}},\ }\href@noop {} {\bibfield  {journal}
  {\bibinfo  {journal} {J. Phys. Chem. C}\ }\textbf {\bibinfo {volume} {113}},\
  \bibinfo {pages} {14798} (\bibinfo {year} {2009})}\BibitemShut {NoStop}%
\bibitem [{\citenamefont {Luo}\ \emph {et~al.}(2009)\citenamefont {Luo},
  \citenamefont {Zhu}, \citenamefont {Zhang}, \citenamefont {Liang},
  \citenamefont {Rao}, \citenamefont {Li},\ and\ \citenamefont
  {Du}}]{luo2009spin}%
  \BibitemOpen
  \bibfield  {author} {\bibinfo {author} {\bibfnamefont {J.}~\bibnamefont
  {Luo}}, \bibinfo {author} {\bibfnamefont {H.}~\bibnamefont {Zhu}}, \bibinfo
  {author} {\bibfnamefont {F.}~\bibnamefont {Zhang}}, \bibinfo {author}
  {\bibfnamefont {J.}~\bibnamefont {Liang}}, \bibinfo {author} {\bibfnamefont
  {G.}~\bibnamefont {Rao}}, \bibinfo {author} {\bibfnamefont {J.}~\bibnamefont
  {Li}}, \ and\ \bibinfo {author} {\bibfnamefont {Z.}~\bibnamefont {Du}},\
  }\href@noop {} {\bibfield  {journal} {\bibinfo  {journal} {J. App. Phys.}\
  }\textbf {\bibinfo {volume} {105}},\ \bibinfo {pages} {093925} (\bibinfo
  {year} {2009})}\BibitemShut {NoStop}%
\bibitem [{\citenamefont {Luo}\ \emph {et~al.}(2010)\citenamefont {Luo},
  \citenamefont {Zhu}, \citenamefont {Liang}, \citenamefont {Rao},
  \citenamefont {Li},\ and\ \citenamefont {Du}}]{luo2010tuning}%
  \BibitemOpen
  \bibfield  {author} {\bibinfo {author} {\bibfnamefont {J.}~\bibnamefont
  {Luo}}, \bibinfo {author} {\bibfnamefont {H.}~\bibnamefont {Zhu}}, \bibinfo
  {author} {\bibfnamefont {J.}~\bibnamefont {Liang}}, \bibinfo {author}
  {\bibfnamefont {G.}~\bibnamefont {Rao}}, \bibinfo {author} {\bibfnamefont
  {J.}~\bibnamefont {Li}}, \ and\ \bibinfo {author} {\bibfnamefont
  {Z.}~\bibnamefont {Du}},\ }\href@noop {} {\bibfield  {journal} {\bibinfo
  {journal} {J. Phys. Chem. C}\ }\textbf {\bibinfo {volume} {114}},\ \bibinfo
  {pages} {8782} (\bibinfo {year} {2010})}\BibitemShut {NoStop}%
\bibitem [{\citenamefont {Yamamoto}\ \emph {et~al.}(1974)\citenamefont
  {Yamamoto}, \citenamefont {Endo}, \citenamefont {Shimada},\ and\
  \citenamefont {Takada}}]{Yamamoto}%
  \BibitemOpen
  \bibfield  {author} {\bibinfo {author} {\bibfnamefont {N.}~\bibnamefont
  {Yamamoto}}, \bibinfo {author} {\bibfnamefont {T.}~\bibnamefont {Endo}},
  \bibinfo {author} {\bibfnamefont {M.}~\bibnamefont {Shimada}}, \ and\
  \bibinfo {author} {\bibfnamefont {T.}~\bibnamefont {Takada}},\ }\href@noop {}
  {\bibfield  {journal} {\bibinfo  {journal} {Jpn. J. Appl. Phys}\ }\textbf
  {\bibinfo {volume} {13}},\ \bibinfo {pages} {723} (\bibinfo {year}
  {1974})}\BibitemShut {NoStop}%
\bibitem [{\citenamefont {Lan}\ \emph {et~al.}(2011)\citenamefont {Lan},
  \citenamefont {Gong}, \citenamefont {Liu},\ and\ \citenamefont
  {Yang}}]{langong}%
  \BibitemOpen
  \bibfield  {author} {\bibinfo {author} {\bibfnamefont {C.}~\bibnamefont
  {Lan}}, \bibinfo {author} {\bibfnamefont {J.}~\bibnamefont {Gong}}, \bibinfo
  {author} {\bibfnamefont {S.}~\bibnamefont {Liu}}, \ and\ \bibinfo {author}
  {\bibfnamefont {S.}~\bibnamefont {Yang}},\ }\href@noop {} {\bibfield
  {journal} {\bibinfo  {journal} {Nanoscale Res. Lett.}\ }\textbf {\bibinfo
  {volume} {6}},\ \bibinfo {pages} {133} (\bibinfo {year} {2011})}\BibitemShut
  {NoStop}%
\bibitem [{\citenamefont {Thackeray}(1997)}]{thackeray}%
  \BibitemOpen
  \bibfield  {author} {\bibinfo {author} {\bibfnamefont {M.~M.}\ \bibnamefont
  {Thackeray}},\ }\href@noop {} {\bibfield  {journal} {\bibinfo  {journal}
  {Prog. Solid. St. Chem.}\ }\textbf {\bibinfo {volume} {25}},\ \bibinfo
  {pages} {1} (\bibinfo {year} {1997})}\BibitemShut {NoStop}%
\bibitem [{\citenamefont {Johnson}\ \emph {et~al.}(1997)\citenamefont
  {Johnson}, \citenamefont {Mansuetto}, \citenamefont {Thackeray},
  \citenamefont {Shao-Horn},\ and\ \citenamefont {Hackney}}]{thackeray2}%
  \BibitemOpen
  \bibfield  {author} {\bibinfo {author} {\bibfnamefont {C.~S.}\ \bibnamefont
  {Johnson}}, \bibinfo {author} {\bibfnamefont {M.~F.}\ \bibnamefont
  {Mansuetto}}, \bibinfo {author} {\bibfnamefont {M.~M.}\ \bibnamefont
  {Thackeray}}, \bibinfo {author} {\bibfnamefont {Y.}~\bibnamefont
  {Shao-Horn}}, \ and\ \bibinfo {author} {\bibfnamefont {S.~A.}\ \bibnamefont
  {Hackney}},\ }\href@noop {} {\bibfield  {journal} {\bibinfo  {journal} {J.
  Electrochem. Soc.}\ }\textbf {\bibinfo {volume} {144}},\ \bibinfo {pages}
  {2279} (\bibinfo {year} {1997})}\BibitemShut {NoStop}%
\bibitem [{\citenamefont {Shen}\ \emph {et~al.}(1993)\citenamefont {Shen},
  \citenamefont {Zerger}, \citenamefont {DeGuzman}, \citenamefont {Suib},
  \citenamefont {McCurdy}, \citenamefont {Potter},\ and\ \citenamefont
  {O'Young}}]{shenzerger}%
  \BibitemOpen
  \bibfield  {author} {\bibinfo {author} {\bibfnamefont {Y.~F.}\ \bibnamefont
  {Shen}}, \bibinfo {author} {\bibfnamefont {R.~P.}\ \bibnamefont {Zerger}},
  \bibinfo {author} {\bibfnamefont {R.~N.}\ \bibnamefont {DeGuzman}}, \bibinfo
  {author} {\bibfnamefont {S.~L.}\ \bibnamefont {Suib}}, \bibinfo {author}
  {\bibfnamefont {L.}~\bibnamefont {McCurdy}}, \bibinfo {author} {\bibfnamefont
  {D.~I.}\ \bibnamefont {Potter}}, \ and\ \bibinfo {author} {\bibfnamefont
  {C.~L.}\ \bibnamefont {O'Young}},\ }\href@noop {} {\bibfield  {journal}
  {\bibinfo  {journal} {Science}\ }\textbf {\bibinfo {volume} {260}},\ \bibinfo
  {pages} {511} (\bibinfo {year} {1993})}\BibitemShut {NoStop}%
\bibitem [{\citenamefont {Devaraj}\ and\ \citenamefont
  {Munichandraiah}(2008)}]{devaraj}%
  \BibitemOpen
  \bibfield  {author} {\bibinfo {author} {\bibfnamefont {S.}~\bibnamefont
  {Devaraj}}\ and\ \bibinfo {author} {\bibfnamefont {N.}~\bibnamefont
  {Munichandraiah}},\ }\href@noop {} {\bibfield  {journal} {\bibinfo  {journal}
  {J. Phys. Chem. C}\ }\textbf {\bibinfo {volume} {112}},\ \bibinfo {pages}
  {4406} (\bibinfo {year} {2008})}\BibitemShut {NoStop}%
\bibitem [{\citenamefont {Kanamori}(1959)}]{kanamori}%
  \BibitemOpen
  \bibfield  {author} {\bibinfo {author} {\bibfnamefont {J.}~\bibnamefont
  {Kanamori}},\ }\href@noop {} {\bibfield  {journal} {\bibinfo  {journal} {J.
  Phys. Chem. Solids}\ }\textbf {\bibinfo {volume} {10}},\ \bibinfo {pages}
  {87} (\bibinfo {year} {1959})}\BibitemShut {NoStop}%
\bibitem [{\citenamefont {Goodenough}(1955)}]{good1}%
  \BibitemOpen
  \bibfield  {author} {\bibinfo {author} {\bibfnamefont {J.~B.}\ \bibnamefont
  {Goodenough}},\ }\href@noop {} {\bibfield  {journal} {\bibinfo  {journal}
  {Phys. Rev.}\ }\textbf {\bibinfo {volume} {100}},\ \bibinfo {pages} {564}
  (\bibinfo {year} {1955})}\BibitemShut {NoStop}%
\bibitem [{\citenamefont {Goodenough}(1958)}]{good2}%
  \BibitemOpen
  \bibfield  {author} {\bibinfo {author} {\bibfnamefont {J.~B.}\ \bibnamefont
  {Goodenough}},\ }\href@noop {} {\bibfield  {journal} {\bibinfo  {journal} {J.
  Phys. Chem. Solids}\ }\textbf {\bibinfo {volume} {6}},\ \bibinfo {pages}
  {287} (\bibinfo {year} {1958})}\BibitemShut {NoStop}%
\bibitem [{\citenamefont {Kajimoto}\ \emph {et~al.}(2002)\citenamefont
  {Kajimoto}, \citenamefont {Mochizuki}, \citenamefont {Yoshizawa},
  \citenamefont {Kimura},\ and\ \citenamefont {Tokura}}]{C_AFM_Kajimoto}%
  \BibitemOpen
  \bibfield  {author} {\bibinfo {author} {\bibfnamefont {R.}~\bibnamefont
  {Kajimoto}}, \bibinfo {author} {\bibfnamefont {H.}~\bibnamefont {Mochizuki}},
  \bibinfo {author} {\bibfnamefont {H.}~\bibnamefont {Yoshizawa}}, \bibinfo
  {author} {\bibfnamefont {T.}~\bibnamefont {Kimura}}, \ and\ \bibinfo {author}
  {\bibfnamefont {Y.}~\bibnamefont {Tokura}},\ }\href@noop {} {\bibfield
  {journal} {\bibinfo  {journal} {Physica B}\ }\textbf {\bibinfo {volume}
  {312}},\ \bibinfo {pages} {760–762} (\bibinfo {year} {2002})}\BibitemShut
  {NoStop}%
\bibitem [{\citenamefont {Metropolis}\ \emph {et~al.}(1953)\citenamefont
  {Metropolis}, \citenamefont {Rosenbluth}, \citenamefont {Rosenbluth},\ and\
  \citenamefont {Teller}}]{MC}%
  \BibitemOpen
  \bibfield  {author} {\bibinfo {author} {\bibfnamefont {N.}~\bibnamefont
  {Metropolis}}, \bibinfo {author} {\bibfnamefont {A.~W.}\ \bibnamefont
  {Rosenbluth}}, \bibinfo {author} {\bibfnamefont {M.~N.}\ \bibnamefont
  {Rosenbluth}}, \ and\ \bibinfo {author} {\bibfnamefont {A.~H.}\ \bibnamefont
  {Teller}},\ }\href@noop {} {\bibfield  {journal} {\bibinfo  {journal} {J.
  Chem. Phys.}\ }\textbf {\bibinfo {volume} {21}},\ \bibinfo {pages} {1087}
  (\bibinfo {year} {1953})}\BibitemShut {NoStop}%
\bibitem [{\citenamefont {Villain}(1977)}]{villain1977spin}%
  \BibitemOpen
  \bibfield  {author} {\bibinfo {author} {\bibfnamefont {J.}~\bibnamefont
  {Villain}},\ }\href {http://stacks.iop.org/0022-3719/10/i=10/a=014}
  {\bibfield  {journal} {\bibinfo  {journal} {J. Phys. C: Solid State Phys.}\
  }\textbf {\bibinfo {volume} {10}},\ \bibinfo {pages} {1717} (\bibinfo {year}
  {1977})}\BibitemShut {NoStop}%
\bibitem [{\citenamefont {Shen}\ \emph {et~al.}(2005)\citenamefont {Shen},
  \citenamefont {Ding}, \citenamefont {Liu}, \citenamefont {Han}, \citenamefont
  {Budnick}, \citenamefont {Hines},\ and\ \citenamefont {Suib}}]{Shen}%
  \BibitemOpen
  \bibfield  {author} {\bibinfo {author} {\bibfnamefont {F.}~\bibnamefont
  {Shen}}, \bibinfo {author} {\bibfnamefont {Y.}~\bibnamefont {Ding}}, \bibinfo
  {author} {\bibfnamefont {J.}~\bibnamefont {Liu}}, \bibinfo {author}
  {\bibfnamefont {Z.~H.}\ \bibnamefont {Han}}, \bibinfo {author} {\bibfnamefont
  {J.}~\bibnamefont {Budnick}}, \bibinfo {author} {\bibfnamefont {W.~A.}\
  \bibnamefont {Hines}}, \ and\ \bibinfo {author} {\bibfnamefont {S.~L.}\
  \bibnamefont {Suib}},\ }\href@noop {} {\bibfield  {journal} {\bibinfo
  {journal} {J. Am. Chem. Soc.}\ }\textbf {\bibinfo {volume} {127}},\ \bibinfo
  {pages} {6166} (\bibinfo {year} {2005})}\BibitemShut {NoStop}%
\bibitem [{\citenamefont {Marinari}\ and\ \citenamefont
  {Parisi}(1992)}]{marinari1992simulated}%
  \BibitemOpen
  \bibfield  {author} {\bibinfo {author} {\bibfnamefont {E.}~\bibnamefont
  {Marinari}}\ and\ \bibinfo {author} {\bibfnamefont {G.}~\bibnamefont
  {Parisi}},\ }\href {http://stacks.iop.org/0295-5075/19/i=6/a=002} {\bibfield
  {journal} {\bibinfo  {journal} {Europhys. Lett.}\ }\textbf {\bibinfo {volume}
  {19}},\ \bibinfo {pages} {451} (\bibinfo {year} {1992})}\BibitemShut
  {NoStop}%
\bibitem [{\citenamefont {Hukushima}\ and\ \citenamefont
  {Nemoto}(1996)}]{hukushima1996exchange}%
  \BibitemOpen
  \bibfield  {author} {\bibinfo {author} {\bibfnamefont {K.}~\bibnamefont
  {Hukushima}}\ and\ \bibinfo {author} {\bibfnamefont {K.}~\bibnamefont
  {Nemoto}},\ }\href {\doibase 10.1143/JPSJ.65.1604} {\bibfield  {journal}
  {\bibinfo  {journal} {J. Phys. Soc. Jpn.}\ }\textbf {\bibinfo {volume}
  {65}},\ \bibinfo {pages} {1604} (\bibinfo {year} {1996})}\BibitemShut
  {NoStop}%
\bibitem [{\citenamefont {Hartmann}\ and\ \citenamefont
  {Rieger}(2002)}]{hartmann2002optimization}%
  \BibitemOpen
  \bibfield  {author} {\bibinfo {author} {\bibfnamefont {A.}~\bibnamefont
  {Hartmann}}\ and\ \bibinfo {author} {\bibfnamefont {H.}~\bibnamefont
  {Rieger}},\ }\href@noop {} {\emph {\bibinfo {title} {Optimization algorithms
  in physics}}}\ (\bibinfo  {publisher} {Berlin: Wiley-VCH},\ \bibinfo {year}
  {2002})\BibitemShut {NoStop}%
\bibitem [{\citenamefont {Binder}(1981)}]{Binder}%
  \BibitemOpen
  \bibfield  {author} {\bibinfo {author} {\bibfnamefont {K.}~\bibnamefont
  {Binder}},\ }\href@noop {} {\bibfield  {journal} {\bibinfo  {journal} {Z.
  Phys. B}\ }\textbf {\bibinfo {volume} {43}},\ \bibinfo {pages} {119}
  (\bibinfo {year} {1981})}\BibitemShut {NoStop}%
\bibitem [{\citenamefont {Bhatt}\ and\ \citenamefont {Young}(1988)}]{Bhatt}%
  \BibitemOpen
  \bibfield  {author} {\bibinfo {author} {\bibfnamefont {R.~N.}\ \bibnamefont
  {Bhatt}}\ and\ \bibinfo {author} {\bibfnamefont {A.~P.}\ \bibnamefont
  {Young}},\ }\href {\doibase 10.1103/PhysRevB.37.5606} {\bibfield  {journal}
  {\bibinfo  {journal} {Phys. Rev. B}\ }\textbf {\bibinfo {volume} {37}},\
  \bibinfo {pages} {5606} (\bibinfo {year} {1988})}\BibitemShut {NoStop}%
\bibitem [{\citenamefont {Kawashima}\ and\ \citenamefont
  {Young}(1996)}]{Kawashima}%
  \BibitemOpen
  \bibfield  {author} {\bibinfo {author} {\bibfnamefont {N.}~\bibnamefont
  {Kawashima}}\ and\ \bibinfo {author} {\bibfnamefont {A.~P.}\ \bibnamefont
  {Young}},\ }\href {\doibase 10.1103/PhysRevB.53.R484} {\bibfield  {journal}
  {\bibinfo  {journal} {Phys. Rev. B}\ }\textbf {\bibinfo {volume} {53}},\
  \bibinfo {pages} {R484} (\bibinfo {year} {1996})}\BibitemShut {NoStop}%
\bibitem [{\citenamefont {Ballesteros}\ \emph {et~al.}(2000)\citenamefont
  {Ballesteros}, \citenamefont {Cruz}, \citenamefont {Fernandez}, \citenamefont
  {Martin-Mayor}, \citenamefont {Pech}, \citenamefont {Ruiz-Lorenzo},
  \citenamefont {Tarancon}, \citenamefont {Tellez}, \citenamefont {Ullod},\
  and\ \citenamefont {Ungil}}]{Ballesteros}%
  \BibitemOpen
  \bibfield  {author} {\bibinfo {author} {\bibfnamefont {H.~G.}\ \bibnamefont
  {Ballesteros}}, \bibinfo {author} {\bibfnamefont {A.}~\bibnamefont {Cruz}},
  \bibinfo {author} {\bibfnamefont {L.~A.}\ \bibnamefont {Fernandez}}, \bibinfo
  {author} {\bibfnamefont {V.}~\bibnamefont {Martin-Mayor}}, \bibinfo {author}
  {\bibfnamefont {J.}~\bibnamefont {Pech}}, \bibinfo {author} {\bibfnamefont
  {J.~J.}\ \bibnamefont {Ruiz-Lorenzo}}, \bibinfo {author} {\bibfnamefont
  {A.}~\bibnamefont {Tarancon}}, \bibinfo {author} {\bibfnamefont
  {P.}~\bibnamefont {Tellez}}, \bibinfo {author} {\bibfnamefont {C.~L.}\
  \bibnamefont {Ullod}}, \ and\ \bibinfo {author} {\bibfnamefont
  {C.}~\bibnamefont {Ungil}},\ }\href {\doibase 10.1103/PhysRevB.62.14237}
  {\bibfield  {journal} {\bibinfo  {journal} {Phys. Rev. B}\ }\textbf {\bibinfo
  {volume} {62}},\ \bibinfo {pages} {14237} (\bibinfo {year}
  {2000})}\BibitemShut {NoStop}%
\bibitem [{\citenamefont {Herrero}(2009)}]{Herrero}%
  \BibitemOpen
  \bibfield  {author} {\bibinfo {author} {\bibfnamefont {C.~P.}\ \bibnamefont
  {Herrero}},\ }\href@noop {} {\bibfield  {journal} {\bibinfo  {journal} {Eur.
  Phys. J. B}\ }\textbf {\bibinfo {volume} {70}},\ \bibinfo {pages} {435}
  (\bibinfo {year} {2009})}\BibitemShut {NoStop}%
\bibitem [{\citenamefont {J\"org}(2006)}]{Thomas}%
  \BibitemOpen
  \bibfield  {author} {\bibinfo {author} {\bibfnamefont {T.}~\bibnamefont
  {J\"org}},\ }\href {\doibase 10.1103/PhysRevB.73.224431} {\bibfield
  {journal} {\bibinfo  {journal} {Phys. Rev. B}\ }\textbf {\bibinfo {volume}
  {73}},\ \bibinfo {pages} {224431} (\bibinfo {year} {2006})}\BibitemShut
  {NoStop}%
\bibitem [{\citenamefont {Hasenbusch}\ \emph {et~al.}(2008)\citenamefont
  {Hasenbusch}, \citenamefont {Pelissetto},\ and\ \citenamefont
  {Vicari}}]{hasenbusch2008critical}%
  \BibitemOpen
  \bibfield  {author} {\bibinfo {author} {\bibfnamefont {M.}~\bibnamefont
  {Hasenbusch}}, \bibinfo {author} {\bibfnamefont {A.}~\bibnamefont
  {Pelissetto}}, \ and\ \bibinfo {author} {\bibfnamefont {E.}~\bibnamefont
  {Vicari}},\ }\href {\doibase 10.1103/PhysRevB.78.214205} {\bibfield
  {journal} {\bibinfo  {journal} {Phys. Rev. B}\ }\textbf {\bibinfo {volume}
  {78}},\ \bibinfo {pages} {214205} (\bibinfo {year} {2008})}\BibitemShut
  {NoStop}%
\end{thebibliography}%

\end{document}